\documentstyle[psbox]{WORKSHOP}

\begin{document}

\title{
3D Dynamical Modeling of Wind Accretion in Cyg X-3
}

\author{
Atsuo T. Okazaki$^1$, Christopher M. P. Russell$^{1}$
\\[12pt]  
%
$^1$  Faculty of Engineering, Hokkai-Gakuen University, Toyohira-ku, Sapporo 062-8605, Japan \\
%
{\it E-mail(ATO): okazaki@lst.hokkai-s-u.ac.jp} 
}

\abst{
Cyg X-3 is a high mass X-ray binary consisting of a Wolf-Rayet star and a compact object in a very short orbital period of 4.8~h. The only confirmed microquasar with high energy gamma-ray emission, Cyg X-3 provides a unique opportunity to study the relationship between the accretion power and the power in high energy emission. Because of a compact orbit and a slow Wolf-Rayet wind, the flow structure around the compact object is thought to be strongly affected by the orbital motion, details of which can be obtained only by numerical simulations.  In this paper, we report on the results from 3D hydrodynamic simulations of the wind accretion in Cyg X-3. For simplicity we adopt an anti-gravity-like force that emulates the radiative acceleration consistent with the beta-velocity wind. Due to the rapid orbital motion, the flow around the compact object has large density gradients. As a result, the accretion rate onto the compact object is significantly lower than that of the Bondi-Hoyle-Lyttleton rate. We also calculate the model X-ray light curve. Although it roughly agrees with the observed light curve, more detailed modeling is needed for detailed comparison.
}

\kword{accretion --- stars: binaries --- stars: individual (Cyg X-3) --- X-rays: binaries}

\maketitle
\thispagestyle{empty}

\section{Introduction}

Cyg~X-3 is a unique laboratory for high energy astrophysics.
It is a very bright, high mass X-ray binary consisting of a Wolf-Rayet star and a compact object,
which is most likely a black hole,
in a circular 4.8~h orbit.
At present, it is the only microquasar from which variable GeV gamma-ray emission has been detected.

Cyg~X-3 exhibits a 4.8~h modulation in multi-wavebands, which should provide important clues 
for the accretion/ejection mechnism.
The X-ray modulation is basically explained by the absorption/scattering in the dense Wolf-Rayet wind, 
but the origin of the asymmetry (\lq\lq slow rise, fast decay'') remains an open question 
(Zdziarski et al.\ 2012; Vilhu \& Hannikainen 2013). 


Fast orbital motion and slow acceleration of the wind (due to a lack of hydrogen) means that 
the Coriolis force could play an important role to shape the accretion pattern.
Therefore, constructing a 3D dynamic model of wind accretion, taking account of the orbital motion, 
is important to understand the high energy activity of Cyg X-3.

\section{Numerical Setup}

We use a 3D SPH code to simulate the Wolf-Rayet wind around the compact object of Cyg~X-3. 
The code uses the variable smoothing length and individual time-steps.
The standard values of artificial viscosity parameters are adopted, i.e., $\alpha=1$ and $\beta=2$.
In one simulation shown below, the wind is launched spherically symmetrically
and the optically-thin radiative cooling is taken into account, while in the other simulations,
isothermal wind particles are ejected only in a narrow range of azimuthal and vertical angles 
toward the compact object, in order to optimize the resolution and computational efficiency.
In these simulations, the Wolf-Rayet wind is assumed to have a velocity distribution of the form
\begin{equation}
v = v_\infty (1-r/R_{\mathrm WR})^\beta,
\end{equation}
where $v_\infty$ is the terminal speed of the wind and $R_{\mathrm WR}$ is the radius of the Wolf-Rayet star.
The stellar and wind paraemters adopted are as follows: $M_{\mathrm WR} = 10.3\,M_\odot$, $R_{\mathrm WR} = 6.1\,R_\odot$, 
$M_{\mathrm X} = 2.4\,M_\odot$, $\dot{M}_{\mathrm WR} = 6.5 \times 10^{-6}\,M_\odot\;\mathrm{yr}^{-1}$, and
$v_\infty = 1,700\,\mathrm{km\;s}^{-1}$ (Zdziarski et al.\ 2013).
The binary orbit is set in the $x$-$y$ plane. In what follows, Phase 0 corresponds to the superior conjunction.

\section{Numerical Results}

\begin{figure*}[t]
\centering
\psbox[xsize=13cm]{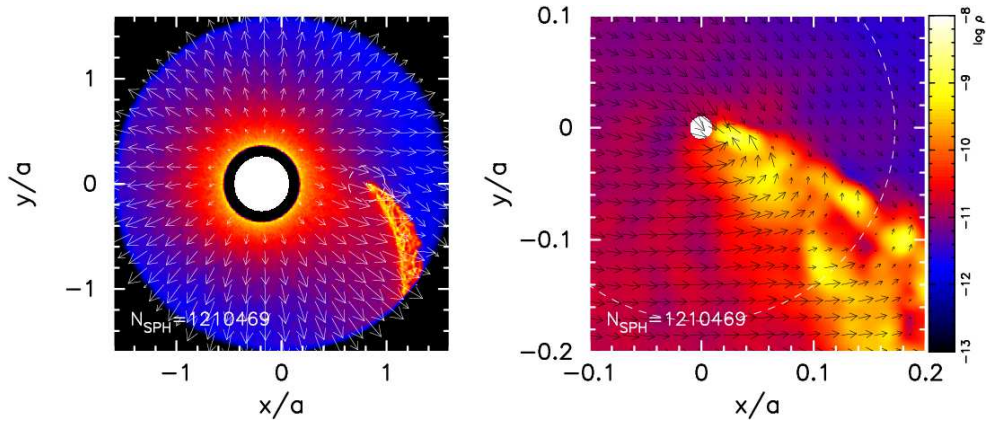}
\caption{ Strucute of a $\beta=2$ Wolf-Rayet wind in the orbital plane. 
For numerical reasons, the wind particles are launched at the radius where 
the wind speed becomes twice the sound speed.
In each plot, the color-scale plot shows the density distribution, 
while arrows denote the velocity vectors.
The dashed circle shows the Bondi-Hoyle-Lyttleton accretion radius. }
\label{fig1}
\end{figure*}

Figure~\ref{fig1} shows the global structure of the $\beta=2$ wind in the orbital plane.
We notice that the Wolf-Rayet wind, which is launched spherically symetrically,
has a large-scale asymmetry around the compact object:
significantly denser gas flow is seen
on the rear side of the compact object than on the front side.
This is because for $\beta = 2$ the wind speed near the compact object is comparable to
the orbital speed of the compact object.

Table~\ref{table1} compares the accretion rates in isothermal simulations for
$\beta=$1, 2, and 3 with the correspronding Bondi-Hoyle-Lyttleton (BHL) accretion rate.
The simulated accretion rate is comparable with the BHL accretion rate for $\beta=1$,
but is significantly lower for $\beta \ge 2$.
As a result, the increase in the simulated accretion rate with increasing $\beta$
is weaker than in the BHL rate.
This is because the flow around the compact object has a larger density gradient for
larger $\beta$, while the BHL accretion rate is for a flow with uniform density distribution.

\begin{table}[!ht]
\caption{Comaparison between the simulated and BHL accretion rates.}
\begin{center}
\begin{tabular}{cccc} \hline\hline\\[-6pt]
$\beta$ & simulation & simulated $\dot{M}$ & BHL $\dot{M}$ \\
&& ($10^{18}\,\mathrm{g\;s}^{-1}$) & ($10^{18}\,\mathrm{g\;s}^{-1}$) \\  \hline

1 & isothermal & 1.6 & 1.7 \\
2 & isothermal & 2.8 & 4.1 \\
2 & radiative cooling & 2.5 & 4.1 \\
3 & isothermal & 4.0 & 8.6 \\  \hline
\end{tabular}
\end{center}
\label{table1}
\end{table}

\begin{figure}[t]
\centering
\psbox[xsize=9cm]{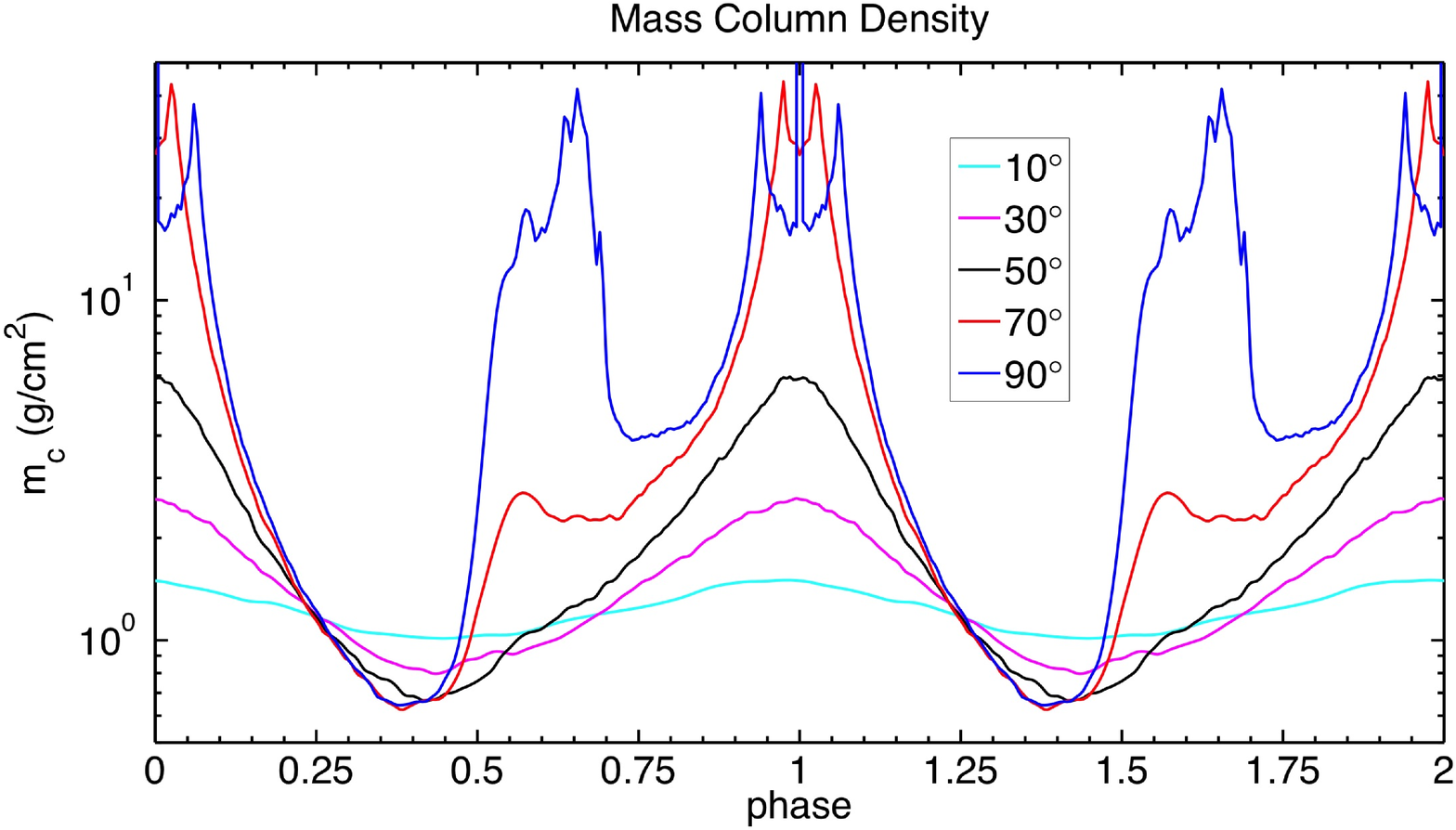}\\
\psbox[xsize=9cm]{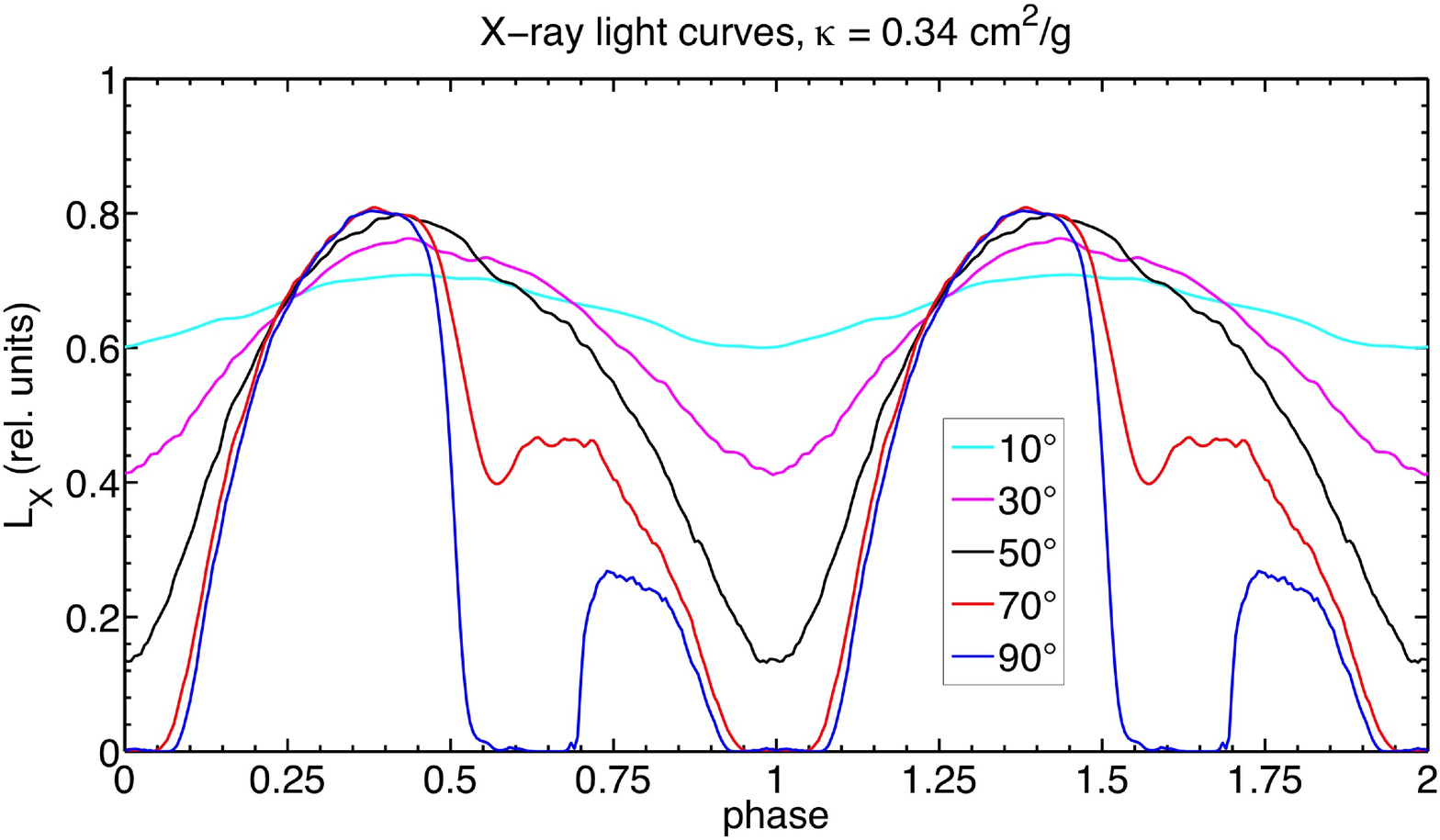}
\caption{Simulated modulation in mass column and 
resulting X-ray light curves for a $\beta=2$ Wolf-Rayet wind with radiative cooling.
The degrees in the legends are inclination angles. }
\label{fig2}
\end{figure}

Finally, Figure~\ref{fig2} shows simulated modulation in mass column and 
resulting X-ray light curves for a $\beta=2$ wind with optically thin, radiative cooling.
Moderate modulation is seen for small to intermediate inclination angles, 
while for high inclination angles, 
absorption/scattering by the accretion wake (phase 0.75) 
becomes remarkable.
Unfortunately, the current model light curves do not reproduce 
the observed light-curve asymmetry.
They are, however, computed for the first time based on the 3D dynamical simulation, where
the effect of the orbital motion is also taken into account.
We will continue our effort to better model the wind structure and absorption/scattering
by the wind.

\vspace{1pc}
\noindent We thank Shunji Kitamoto for helpful comments on the X-ray behavior of Cyg X-3.
This work was partially supported by the JSPS Grants-in-Aid for Scientific Research (C) 
(24540235).

\section*{References}

\re
Zdziarski, A.A. et al. 2012 MNRAS., 426, 1031

\re
Zdziarski, A.A. et al. 2013 MNRASL., 429, L104

\re
Vilhu, O. \& Hannikainen, D.C. 2013 A\&A., 550, A48

\label{last}

\end{document}